\documentclass[fleqn,10pt]{wlpeerj}
\usepackage{array}
\usepackage{longtable}
\usepackage{tabularx}
\usepackage{float}
\usepackage{longtable}
\usepackage{graphicx}

\title{Dependable Artificial Intelligence with Reliability and Security (DAIReS): A Unified Syndrome Decoding Approach for Hallucination and Backdoor Trigger Detection }

\makeatletter
\let\if@peerjlineno\iffalse
\makeatother

\author[1]{Hema Karnam Surendrababu}
\author[2]{Nithin Nagaraj}
\affil[1]{School of Conflict and Security Studies, National Institute of Advanced Studies, Indian Institute of Science Campus, Bengaluru-India.}
\affil[2]{Complex Systems Programme, National Institute of Advanced Studies, Indian Institute of Science Campus, Bengaluru-India}
\corrauthor[1]{Hema Karnam Surendrababu}{karnamshema@gmail.com}

\begin{abstract}

Machine Learning (ML) models, including Large Language Models (LLMs), are characterized by a range of system-level attributes such as security and reliability. Recent studies have demonstrated that ML models are vulnerable to multiple forms of security violations, among which backdoor data-poisoning attacks represent a particularly insidious threat, enabling unauthorized model behavior and systematic misclassification. In parallel, deficiencies in model reliability can manifest as hallucinations in LLMs, leading to unpredictable outputs and substantial risks for end users. In this work on Dependable Artificial Intelligence with Reliability and Security (DAIReS), we propose a novel unified approach based on \textit{Syndrome Decoding} for the detection of both security and reliability violations in learning-based systems. Specifically, we adapt the syndrome decoding approach to the NLP sentence-embedding space, enabling the discrimination of poisoned and non-poisoned samples within ML training datasets. Additionally, the same methodology can effectively detect hallucinated content due to self referential meta explanation tasks in LLMs.
\end{abstract}

\begin{document}

\flushbottom
\maketitle
\thispagestyle{empty}

\section*{Introduction}

Recent years have witnessed a transformative advancement of Artificial Intelligence (AI) from narrow models with limited capabilities to highly capable Large Language Models (LLMs) with enhanced capabilities. Notable capabilities of the LLMs encompass Natural Language Generation, Toxic Content detection, code generation, multimodal interaction, medical diagnostics and financial analysis. As a result, LLMs are no longer confined to low-risk applications but are increasingly integrated into high stake domains such as robotics, cybersecurity, healthcare, finance, public policy and scientific research. This growing integration of LLMs into everyday tasks underscores the paramount significance of ensuring the reliability, security, safety and robustness of LLMs.

Although the deployment of LLMs in the real world continues at a rapid pace, recent research works indicate significant security, safety and reliability vulnerabilities of LLMs ~\citep{real_backdoor} ~\citep{clussman2025}~\citep{NLP_review} ~\citep{maleficnet}  ~\citep{BadNL} raising significant concerns over their continued usage in high stake domains.  These reliability and security vulnerabilities can result in unauthorized model behaviour, model misclassifications, model hallucinations posing major risks to the users of LLMs. In this work we examine two significant system attributes of Machine Learning (ML) models and LLMs – security and reliability. The security attribute of a system can be defined as ensuring that the system does what it is supposed to do and nothing else, whereas the reliability of a system ensures that an LLM continues to deliver the intended functionality for a given period of time ~\citep{avizienis2004basic}. From a security standpoint, we focus on a sophisticated form of data poisoning attacks against ML models, known as backdoor attacks~\citep{poisoningattacks}~\citep{spectralsignatures}. In terms of the reliability attribute, this study’s specific emphasis is on detecting hallucinations in the current state of the art LLMs. 

Amongst the many security violations of LLMs, a sophisticated form of security violation occurs via the backdoor attacks. An adversary mounts a backdoor attack against an ML model by introducing backdoor triggers into a small subset of samples from the training dataset ~\citep{spectralsignatures}~\citep{activationclustering} ~\citep{strip} ~\citep{HKNNIEEE} ~\citep{RAP2021} ~\citep{Exposebackdoors}. Typically, a backdoor trigger can be an embedded pixel pattern inserted into images or perturbations introduced into text data in the training dataset. To ensure that the legitimate functionality of the model is maintained, the adversary crafts the backdoor triggers such that the model produces misclassifications on selected samples that contain the backdoor trigger alone. In other words, the model accuracy continues to meet the set benchmarks with respect to intended functionality, and the model produces unintended behaviour only in the presence of a backdoor trigger. Given this stealthy form of a backdoor attack, it is extremely challenging to detect backdoor attacks via functional testing of ML models. Although LLMs are pretrained, backdoor attacks continue to be relevant in the context of LLMs, as backdoor triggers can be inserted during finetuning, model alignment, and during the instruction tuning stage.

In the context of LLMs, hallucinations can be defined as model generated outputs that are not consistent with provided source context or model outputs that are not grounded in the training data~\citep{ji2023survey} ~\citep{hong2024hallucinations}. While recent research studies proposed diverse metrics to detect hallucinations across varied Natural Language Generation (NLG) tasks, our study addresses a specific form of hallucination that emerges as a consequence of self-referential meta explanation NLG tasks. Self-referential meta explanation prompts typically involve a Large Language Model accessing its own reasoning process while answering user queries. A typical outcome of self-referential prompts is the generation of semantically incoherent output by the LLM. In contrast to current studies that concentrate on metrics that quantify factuality or faithfulness of the generated outputs to the source context, we instead focus on assessing and quantifying the semantic incoherence in LLM generated text. To this end, we provide a quantitative measure for semantic degeneration in LLM outputs.

This work introduces syndrome based decoding, a novel approach that has its foundations in linear block codes and is designed to address both challenges described above - security and reliability aspects of Machine Learning and Large Language Models. Unlike existing approaches~\citep{bang2025hallulens}~\citep{hong2024hallucinations}~\citep{onion}~\citep{strip}~\citep{RAP2021} that rely on different approaches to detect backdoor attacks and hallucination errors, our proposed approach provides a unified framework for defending against various forms of backdoor attacks across diverse domains and reliably detect structural failure modes or Large Language Model Collapse due to self-referential prompts. 

\section*{Related Work}
Although existing research has largely investigated security and reliability attributes of learning models in isolation, we argue that the absence of explicit causal grounding within the models contributes to the manifestation of both the issues. Recognizing this common foundation, we adopt a unified perspective in organizing the related literature. Specifically, we review prior work along two complementary dimensions: (i) defenses that primarily address backdoor attacks, and (ii) approaches that focus on mitigating hallucinations, highlighting both their individual contributions and limitations.

\subsection*{Backdoor Attacks}
The existing defense mechanisms against backdoor attacks address two major classes of threats – a) defenses that mitigate attacks during the pretraining stage ~\citep{spectralsignatures}~\citep{activationclustering}, b) defenses that detect if a model is backdoored during inference stage~\citep{onion} ~\citep{Exposebackdoors} ~\citep{RAP2021}. Defenses during the pretraining stage require training the model on poisonous datasets to detect potential backdoors and retraining the model on the sanitized dataset. On the other hand, defenses during the inference stage assume access to a small subset of non-poisoned samples. Additionally, the defenses are tailormade to specific domains which include NLP, Computer Vision (CV) or tabular datasets from other domains. 

Furthermore, the prevailing defense approaches target a subset of backdoor triggers which include , a) static backdoor trigger where the trigger is a fixed pattern that repeats across multiple training samples~\citep{spectralsignatures} ~\citep{HKNNIEEE} ~\citep{HKETC}, b) paraphrase backdoor trigger where the triggers vary lexically but preserve the same meaning with minor syntactic reordering, synonym substitution etc ~\citep{liu2022piccolo}.   c) semantic or context dependent triggers where the trigger is characterized and activated by semantic meaning or the context instead of a fixed lexical or syntactic form. In other words, a semantic trigger is often activated when a certain conceptual condition is met and remains invariant across multiple training samples, regardless of the semantic equivalence of the sentences.

\subsection*{Hallucinations}
The work in ~\citep{ji2023survey} presents an elaborate survey of hallucinations reported across diverse NLG related tasks and corresponding mitigation methods. The NLG tasks reported in this work include Abstractive or Text Summarization, Dialogue Generation, Data to Text generation, and Neural Machine Translation. However, the hallucinations in the aforementioned tasks are evaluated based on faithfulness of the generated output to the input source context. The metrics for detecting hallucinations include lexical analysis of source and model generated output, entailment probability between the source and the target output potentially quantifying the number of times the model generated output logically follows or contradicts the source content. The automated metrics are further evaluated via a comparative analysis of these metrics against a human annotated gold standard ~\citep{shuster2021retrieval} ~\citep{santhanam2021rome}. In addition, different language models are trained by using general language knowledge and source-conditioned knowledge and hallucinations are detected by identifying discrepancies between their respective outputs ~\citep{filippova-2020-controlled} ~\citep{tian2019sticking}. The work in ~\citep{bang2025hallulens} proposes novel metrics for hallucinations that exclusively stem from gaps in the training data of the LLMs and modelling errors. Specific NLP tasks evaluated in ~\citep{bang2025hallulens} to quantify hallucinations include, long form content generation, short form fact seeking queries, and model queries targeting training data gaps or knowledge gaps. ~\citep{hong2024hallucinations} proposes two metrics – the faithfulness score and factuality score. Faithfulness score is a measure quantifying the LLM’s adherence to the input source content, whereas the factuality score measures the LLM’s capability to generate accurate content based on the world knowledge acquired by the LLM during training. Specific tasks studied in this work include reading comprehension, text summarization and question answering.

In addition to the task specific categorization of the hallucination metrics described above, these metrics can also be classified according to the taxonomy of hallucinations and the specific type of hallucination that they address. ~\citep{ji2023survey} broadly classifies hallucinations as intrinsic and extrinsic, where intrinsic hallucinations are defined as model generated outputs that contradict the input source context, whereas extrinsic hallucinations represent model outputs that cannot be verified or supported or refuted from the source context. Benchmarks that address intrinsic hallucinations have been discussed in ~\citep{bang2025hallulens}~\citep{gu2024anah}~\citep{ming2024faitheval}. Metrics that address extrinsic hallucinations include ~\citep{bang2025hallulens} ~\citep{wei2024measuring} ~\citep{min2023factscore} ~\citep{song2024veriscore}. The work in ~\citep{zhang2025siren} further classifies hallucinations as model generated outputs that are input conflicting, context conflicting and fact conflicting. Metrics that address input conflicting hallucinations, context conflicting hallucination and fact conflicting hallucinations are extensively discussed in ~\citep{zhang2025siren}and ~\citep{lin2022truthfulqa}.

\subsection*{Contribution of Our Work}
Our current work on hallucinations complements the existing body of research by addressing a specific type of hallucination errors that stem from self-referential meta explanation tasks. Self-referential meta explanation prompts can be defined as those prompts that query the model to access its own reasoning process and explain its reasoning to the user while performing a specific NLG task. The model once queried on a self-referential meta explanation task is guaranteed to generate profoundly nonsensical output. Self-referential prompts and the rationale behind the semantically degenerated LLM outputs due to these prompts are elaborated in the subsequent sections.

Despite security and reliability being different dimensions of learning models, we posit that both backdoor attacks and self-referential meta explanation task related hallucinations in modern learning models occur due to an underlying common cause – the model’s limitations in causal inference. Prior backdoor defense mechanisms ~\citep{activationclustering} ~\citep{spectralsignatures} have extensively reported that most ML models learn statistical associations. These statistical associations are in turn exploited by an adversary to introduce malicious correlations between backdoor triggers and model outputs, thereby eliciting unintended and malicious model behaviour. On the other hand, we propose that hallucination errors due to self-referential meta explanation tasks can be attributed to a lack of causal inference structure within the model.  To the best of our knowledge, while backdoor attacks and hallucinations have been extensively surveyed in existing research studies in isolation, a unified framework for simultaneously addressing both challenges with a single methodology is lacking, our proposed approach seeks to bridge this gap. Our contribution is summarized as follows,

\begin{enumerate}
    \item We propose a novel approach that centres on on DAIReS, a unified Syndrome Decoding approach which simultaneously addresses backdoor attack detection and hallucination detection in Large Language Models. To the best of our knowledge, no such unified methodology exists in the current body of literature that addresses two different dimensions of Learning Models – namely security and reliability using the same methodology.
    \item Our approach for backdoor detection successfully evaluates training datasets across multiple domains spanning NLP, tabular data derived from the Social Sciences and geological domains. Data poisoning simulations included testing multiple forms of backdoor triggers including static, and paraphrase sentence triggers.
    \item The proposed approach has successfully tested self-referential meta explanation prompts to objectively detect hallucinations across diverse state of the art LLMs including Claude Sonnet 4.5, ChatGPT 5.2, Gemini 3, Microsoft Copilot, and Perplexity AI.
\end{enumerate}

\section*{Adversarial Threat Model Assumptions and Backdoor Attack Design}
\subsection*{Threat Model}
We employ the threat model established in prior works ~\citep{BadNL} ~\citep{HKNNIEEE}. Under this threat model, the adversary is assumed to possess the capability of injecting backdoor triggers into publicly available training datasets, which are subsequently used by end users to train machine learning models. Specifically, the adversary can perform a dirty-label static or paraphrase sentence backdoor attack by embedding triggers into a small fraction of training samples while simultaneously altering their associated labels to a designated target class. Such threats may originate from a malicious third party involved in data collection, an untrustworthy insider, or a compromised crowdsourcing contributor capable of undermining the integrity of the training corpus. Crucially, the backdoor trigger is designed to remain covert, ensuring that the model’s performance on non-poisoned inputs is not adversely affected in the absence of the trigger.  The defender’s assumptions include having access to the training corpus, and a small set of synthetically generated non-poisonous training data samples.

\subsection*{Backdoor Attack Design}
The experimental setup for designing the backdoor include 6 different datasets. Our evaluation includes a static backdoor trigger experiment conducted across the NLP Stanford Sentiment Treebank (SST-2)~\citep{SST2_data}, Jigsaw Toxicity~\citep{Jigsaw}, Trawling for Trolling Hate speech dataset~\citep{hitkul_2020_3828502}. A static backdoor attack involves inserting a specific text pattern into a small subset of training samples and simultaneously changing the corresponding class label of the samples~\citep{BadNL} ~\citep{HKNNIEEE}. The poisoning ratio is defined as the fraction of poisoned samples in a class of training samples, and for the current experiments a poisoning ratio of $5$\% to $15$\% was employed.  Furthermore, we investigated a paraphrase-based sentence trigger attack. In this class of attack, while no specific trigger pattern is inserted into the training samples, the text data in the training sample is paraphrased by employing a T5 transformer model ~\citep{raffel2020exploring} to preserve semantic meaning of the sentence and simultaneously the class label of the samples is altered to a target class chosen by the adversary. For the tabular datasets, the datasets experimented with include the Forest Cover data ~\citep{covertype_31} and the US Adult Income or Census data ~\citep{adult_ucirepo}. In the case of tabular datasets, we experimented with both the out of bounds numeric trigger, and an inbounds numeric trigger to simulate a static backdoor attack as described in ~\citep{pleiter2023backdoor}. In the case of out of bounds trigger, the trigger value was chosen to be outside the training data distribution, whereas in the case of inbounds trigger the most common feature value was selected as the trigger values.

\section*{Methodology} 
\subsection*{Syndrome Decoding for Backdoor Detection}
The backdoor detection approach formulated in the current work is heavily inspired by the Error Correction Codes (ECC) and adapts syndrome decoding from linear block codes to detect backdoor triggers in training datasets, and additionally for detecting hallucinations in LLMs. An overview of the methodology for backdoor detection is discussed next.

Training data is poisoned by inserting backdoor triggers into a legitimate dataset resulting in malicious training data. To simulate an adversarial backdoor attack, backdoor triggers are inserted into a training dataset that results in a poisoned dataset.  The backdoor attack methods include both a static backdoor attack, and a dynamic paraphrase backdoor attack. The NLP datasets used for the current analysis include SST2, Fake news, Jigsaw Toxicity, Trawling to Trolling, and tabular datasets – Forest cover and U.S. Adult income or Census data from the geological and social sciences domains respectively. The NLP datasets are subsequently transformed into numeric representation via the Sentence BERT paraphrase-mpnet transformer model ~\citep{SBERT}. The backdoors in the training data are detected via the syndrome based decoding approach as elaborated in the subsequent sections.

\subsection*{Preliminaries:  Algebraic Setup for Syndrome Decoding}
Since the proposed method is formulated in terms of syndrome computation, we briefly introduce the required notation and definitions in the context of linear block codes.

\begin{enumerate}
    \item Let $G \in \mathbb{F}_2^{m \times n}$ be the generator matrix that is used to introduce redundancy into the message.
    \item For an input vector $\mathbf{u}$, the encoded representation is given by $\mathbf{c} = \mathbf{u} G$.
    \item Let $H \in \mathbb{F}_2^{m \times n}$ be the denote the parity check matrix that is used to check the validity of a codeword.
    
    \item For a codeword $\mathbf{c}$, the syndrome is defined as $\mathbf{s} = H^{\top}\mathbf{c}$, and for valid codewords $H^{\top}\mathbf{c} = \mathbf{0}$.
    \item The generator matrix $G$ and parity-check matrix $H$ satisfy $G H^{\top} = \mathbf{0}$.
\end{enumerate}

We borrow heavily from linear block coding approach and adapt the syndrome decoding to the NLP sentence embeddings space to distinguish between poisoned and non-poisoned samples in ML training datasets. The $G$ and $H$ matrices in the context of linear block codes operate in $GF_2$ and typically involve binary bit operations. However, a key difference for the current work is to define the generator and parity check matrices in the context of sentence embeddings. To this end, we formulate the generator matrix $G$ to reflect the error/backdoor trigger subspace, and the parity matrix is formulated to reflect the semantic content or meaning of the sentence embeddings in the embedding space.

While semantic meaning can be quantified through the sentence embeddings generated from pretrained models, the key baseline in our approach is to define a projection matrix $G$ that can potentially capture the backdoor trigger or the poisoned subspace that deviates from the semantic subspace. Given that there is an inherent redundancy built into sentences that convey similar meaning, with their embeddings clustered in the sentence embedding space, this redundancy is utilized to build a projection kernel or generator matrix that can factor out the semantic subspace and detect the leftover trigger structure. Accordingly, a generator matrix and parity check matrix are built by the following steps.

\begin{enumerate}
    \item Generate a few synthetic non-poisoned samples from a single class for each dataset. 
    \item Generate paraphrases of the samples generated in step1 using the paraphrase T5 transformer model.
    \item Generate sentence embeddings for the paraphrased sentences in step 2 using SBERT paraphrase-mpnet sentence transformer. Let $F \in \mathbb{R}^{m \times d}$ denote the paraphrased sentence embeddings.
    \item Let $F_{\text{centered}} \in \mathbb{R}^{m \times d}$ denote the mean-centered sentence embedding representation, defined as
    \[
        F_{\text{centered}} = F - \mathbf{1}\mu,
        \quad
        \mu = \frac{1}{m}\sum_{i=1}^{m} F_{i,:},
    \]
    \item Perform Principal Component Analysis (PCA) on the mean centered paraphrased sentence embeddings.
    we compute the sample covariance matrix as
    \[
        \Sigma = \frac{1}{m} F_{\text{centered}}^{\top} F_{\text{centered}}.
    \]
    The eigen-decomposition is given by,
    \[
        \Sigma = U \Lambda U^{\top},
    \]
    where
    \[
        U = [U_1, U_2, \dots, U_l]
    \] contains the orthonormal eigenvectors (principal directions), and
    \[
        \Lambda = \mathrm{diag}(\lambda_1, \lambda_2, \dots, \lambda_l),
        \quad
        \lambda_1 \ge \lambda_2 \ge \cdots \ge \lambda_l \ge 0,
    \]
    are the corresponding eigenvalues.
    \item Select the PCA component with the lowest variance ${U_l}$ as the generator matrix to capture the residual subspace or backdoor trigger structure.
    \item Project the sentence embeddings of the non-poisoned samples spanned by the generator matrix to create an encoded representation of the sentence embeddings. The encoded representation is given by,
    \[
        \operatorname{proj}(F_{\text{centered}})
        = U_l \bigl( U_l^{\top} F_{\text{centered}} \bigr) + \mu
    \]
    \item The parity check matrix to capture the semantic subspace lies in the orthogonal subspace of the generator matrix subspace and is computed as below,
    \[
        (I - U_l U_l^{\top})
    \]
    \item A decoded sentence embedding representation is obtained via the parity check matrix and the syndrome computation is described below,
    \[
        \mathbf{s} = {F_{\text{centered}}} -\operatorname{proj}(F_{\text{centered}})
        = (I - U_l U_l^{\top}){F_{\text{centered}}}.
    \]
    \item The syndrome computed above is utilized as the syndrome template for non-poisoned samples.
\end{enumerate}

A distribution of syndrome magnitudes is formulated as a syndrome template as described in steps 1-10, from the synthetically generated non-poisoned samples. This syndrome template is subsequently employed as a basis for comparative evaluation of syndromes generated from multiple subsets of the training dataset as described next. The training dataset is divided into multiple subsets spanning a few hundred training instances. For each subset, a generator matrix, parity matrix and a corresponding syndrome magnitude is computed as described in steps 3-10.  The syndromes thus computed from each subset  are subsequently compared with the syndrome template to distinguish between poisoned and non-poisoned samples in the training dataset.

For detecting semantic degeneration that stems from self-referential meta explanation prompts in LLMs, the same methodology based on syndrome decoding is employed.  

\subsection*{Design Rationale for the Proposed Syndrome Decoding Approach}
Notably, in the embedding space, the embeddings of the non-poisoned samples are concentrated in close proximity to the semantic subspace, whereas the poisoned samples with the embedded backdoor triggers form a subspace that potentially deviates from the semantically meaningful subspace. Given this and the observation that the generator matrix kernel projects the embeddings onto the lowest variance PCA component (representing the noise/error subspace), it follows that the non-poisoned embeddings when projected using the generator matrix kernel $G$ have very low magnitudes. The low magnitudes are obtained due to the absence of backdoor triggers in the non-poisoned sentence embeddings. In other words, the encoded representation obtained via the generator matrix is indicative of the amount of poison/backdoor trigger present in the sentence embeddings. In the syndrome computation that follows, the encoded representation of non-poisoned embeddings is further projected using the parity matrix $H$. Given that the parity matrix $H$ lies in the orthogonal subspace of $G$ and is defined as the maximum variance PCA component, the projection using the $H$ kernel quantifies the semantic coherence of the sentence embeddings. Additionally, while computing the syndrome magnitude, we employ an inflated encoded representation, to amplify the strength of structural signal while preserving the intrinsic geometry of the embedding space. Uniform scaling boosts the structured signal that is to be detected. This approach is similar to Signal to Noise (SNR) boosting frequently employed in signal processing.

Although in DAIReS, our syndrome based decoding approach borrows algebraic tools from linear block coding, such as parity-check matrices and syndrome computation, unlike the classical syndrome decoding our approach does not constitute an error-correcting code. Instead, the syndrome is used as a structured linear projection and a mapping that can be used to detect or distinguish between non-poisoned and poisoned samples in ML training datasets.

\section*{Results}
An extensive experimental evaluation was conducted across 6 datasets to assess the effectiveness of the proposed syndrome based decoding approach for detection of backdoor attacks in ML training data and semantically degenerated outputs in LLMs. Specifically, 4 NLP datasets and 2 tabular datasets were utilized to evaluate performance on backdoor attack detection, and diverse forms of self-referential meta explanation prompts were employed for detecting semantically degenerated outputs and structural failures of LLMs. This section presents quantitative results across these datasets and LLMs.

\subsection*{Experimental Evaluation - Backdoor Detection}

The datasets analyzed for backdoor detection include the NLP datasets – SST-2 movie reviews, Jigsaw toxicity detection, Trawling for Trolling hate speech data. For the paraphrase sentence trigger attack, trigger phrases were generated from the SST-2 negative class samples and the corresponding class label is altered to an adversary chosen label. The tabular datasets that were evaluated include the Forest Cover dataset from the geological domain and the U.S. Adult Income or Census data from the social sciences domain and both in bounds and out of bounds trigger were simulated for the attack scenario.

\begin{figure}[htbp]
    \centering
    \includegraphics[width=0.95\textwidth]{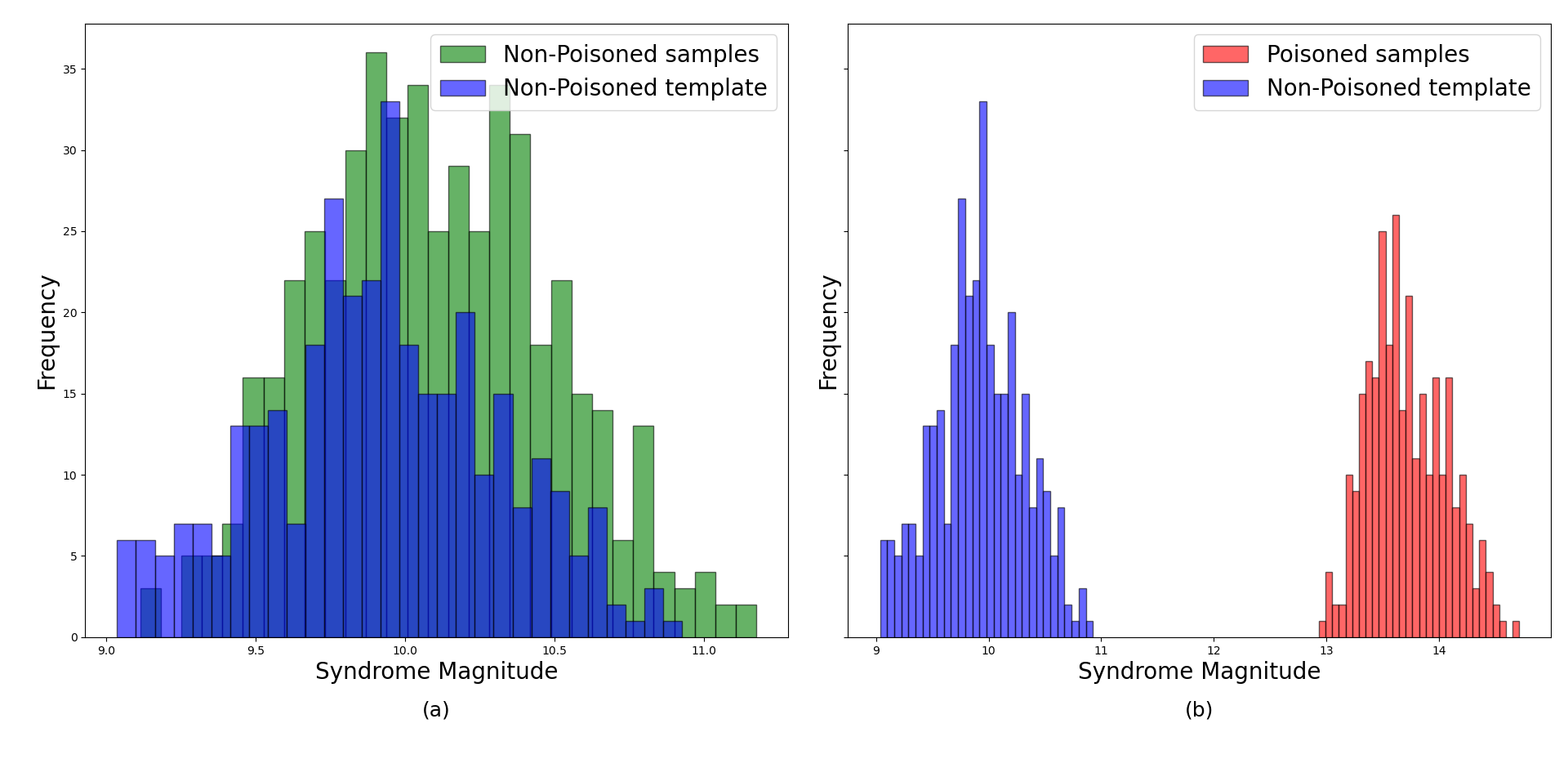}
    \caption{SST-2 dataset, Distribution of Syndrome Magnitudes: (a) non-poisoned training data vs non-poisoned template, (b) poisoned training data vs non-poisoned template.}
    \label{fig:SST2_backdoor_detn}
\end{figure}

\begin{figure}[htbp]
    \centering
    \includegraphics[width=0.95\textwidth]{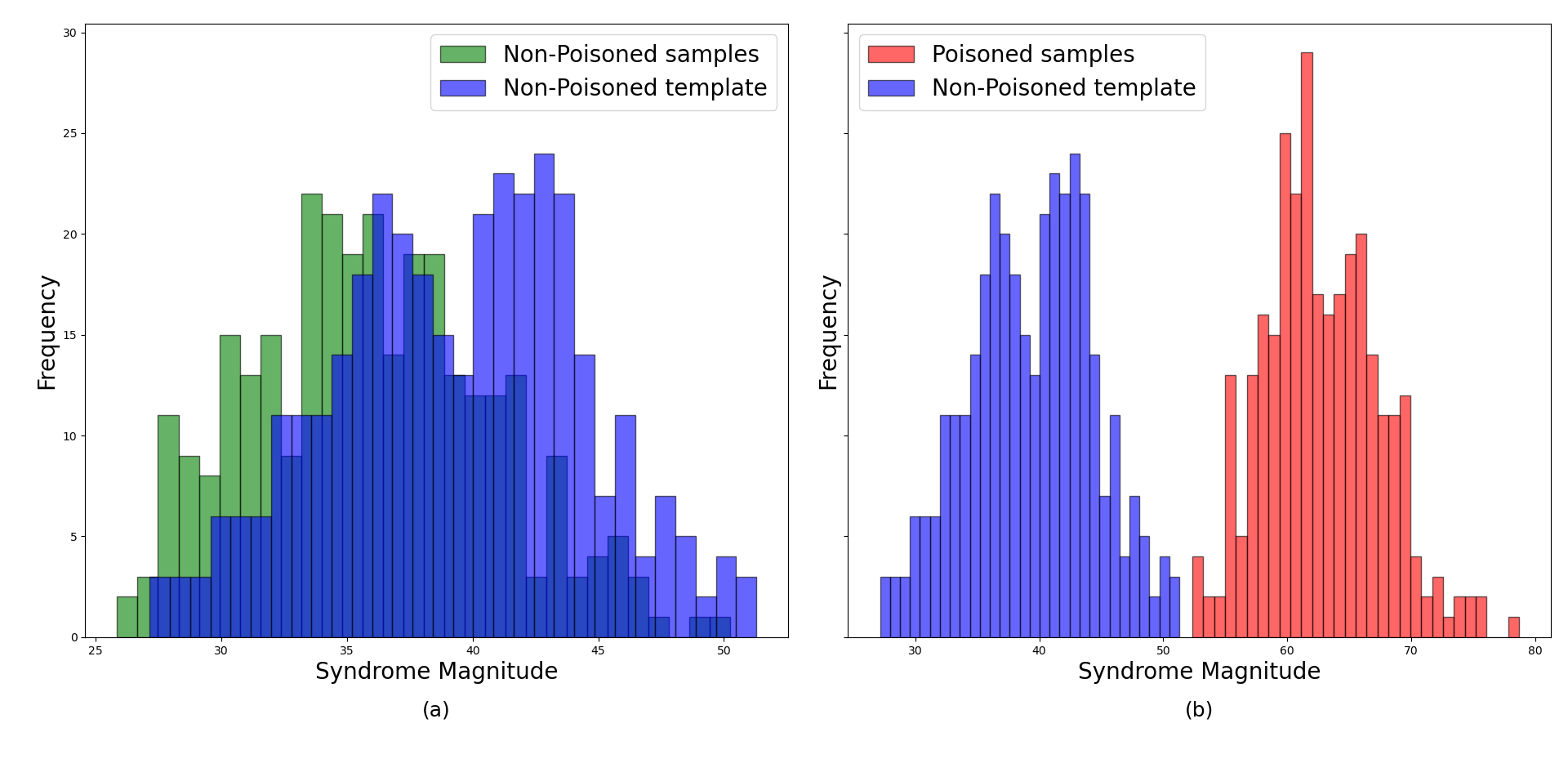}
    \caption{Jigsaw Toxicity dataset, Distribution of Syndrome Magnitudes: (a) non-poisoned training data vs non-poisoned template, (b) poisoned training data vs non-poisoned template.}
    \label{fig:Jigsaw_backdoor_detn}
\end{figure}

\begin{figure}[htbp]
    \centering
    \includegraphics[width=0.95\textwidth]{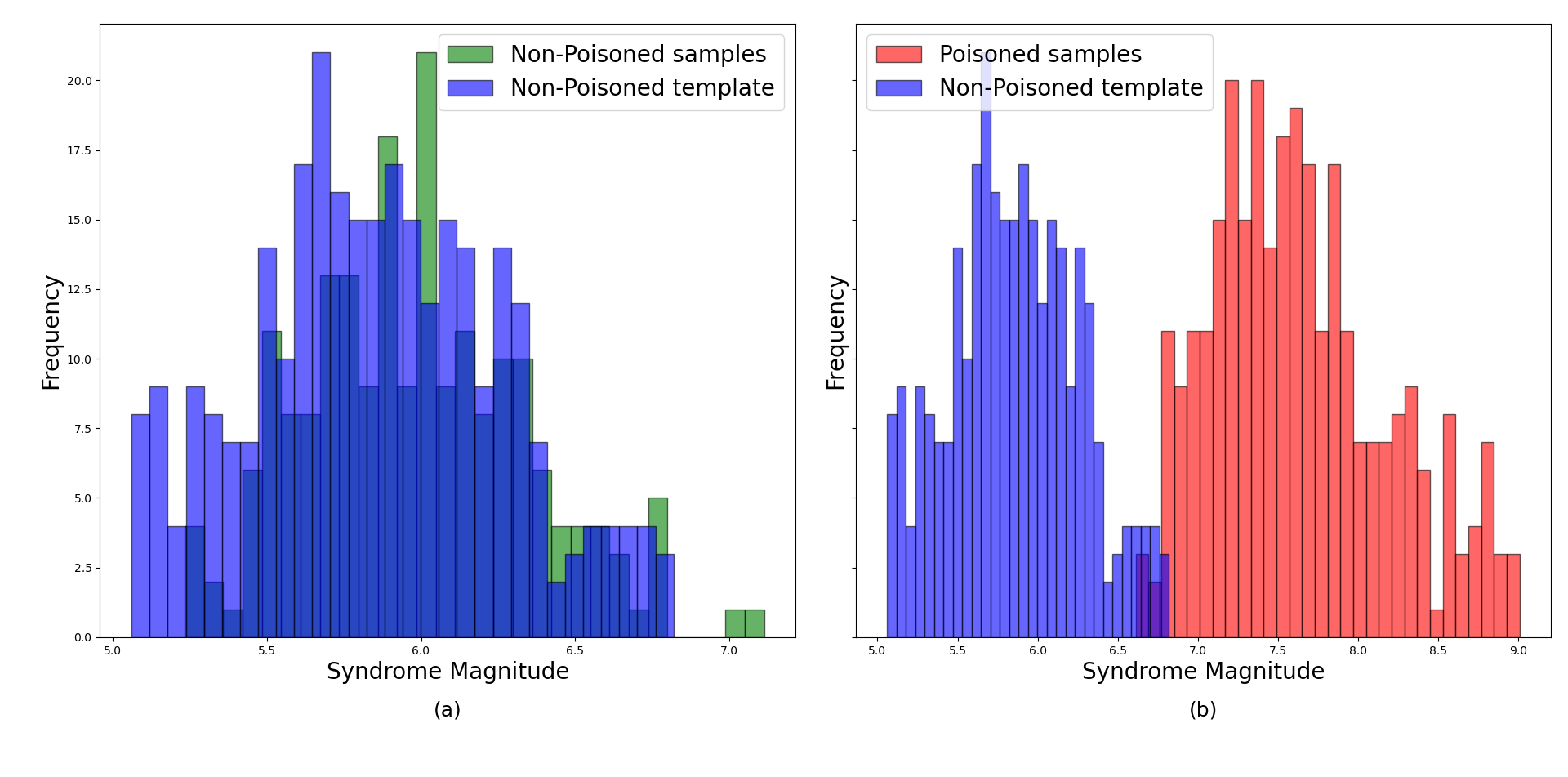}
    \caption{Trawling for Trolling Hate speech dataset dataset, Distribution of Syndrome Magnitudes: (a) non-poisoned training data vs non-poisoned template, (b) poisoned training data vs non-poisoned template.}
    \label{fig:Trolling_backdoor_detn}
\end{figure}

\begin{figure}[htbp]
    \centering
    \includegraphics[width=0.95\textwidth]{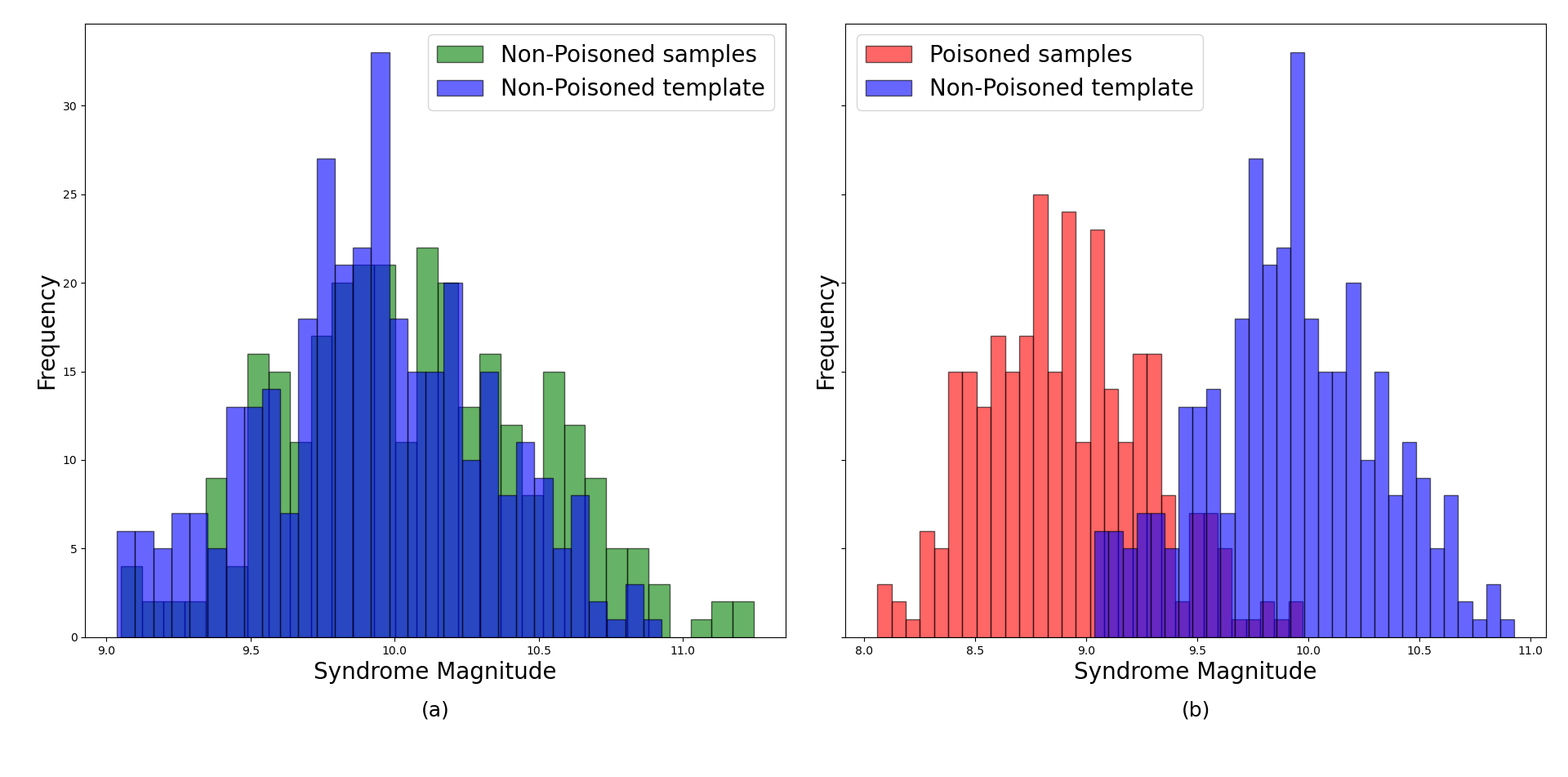}
    \caption{Paraphrase Backdoor Attack, Paraphrases derived from SST-2 dataset, Distribution of Syndrome Magnitudes: (a) non-poisoned training data vs non-poisoned template, (b) poisoned training data vs non-poisoned template.}
    \label{fig:Paraphrases_backdoor_detn}
\end{figure}

\begin{figure}[htbp]
    \centering
    \includegraphics[width=0.95\textwidth]{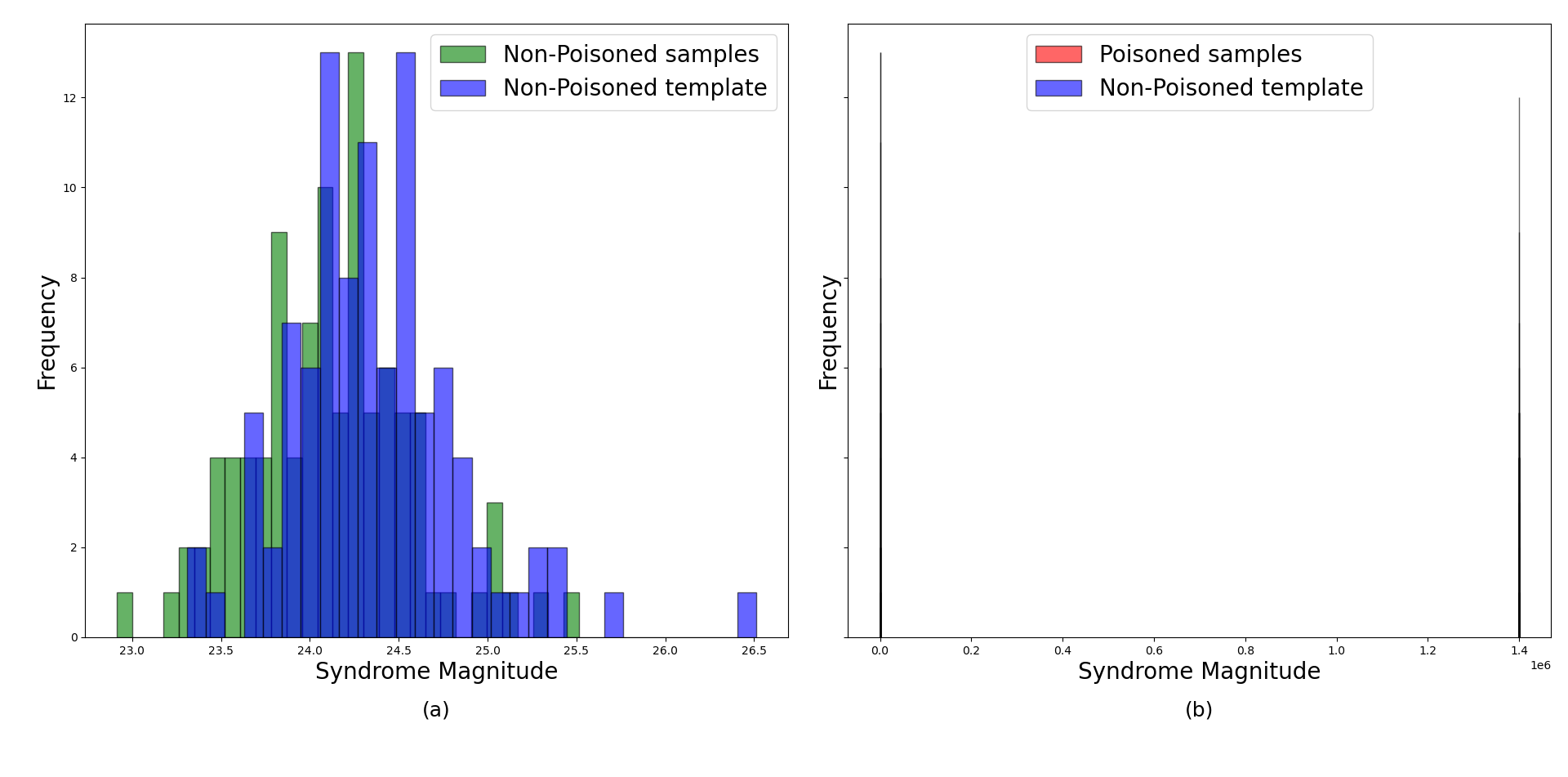}
    \caption{U.S. Adult Income or Census tabular dataset, Distribution of Syndrome Magnitudes: (a) non-poisoned training data vs non-poisoned template, (b) poisoned training data vs non-poisoned template.}
    \label{fig:Adult_backdoor_detn}
\end{figure}

\begin{figure}[htbp]
    \centering
    \includegraphics[width=0.95\textwidth]{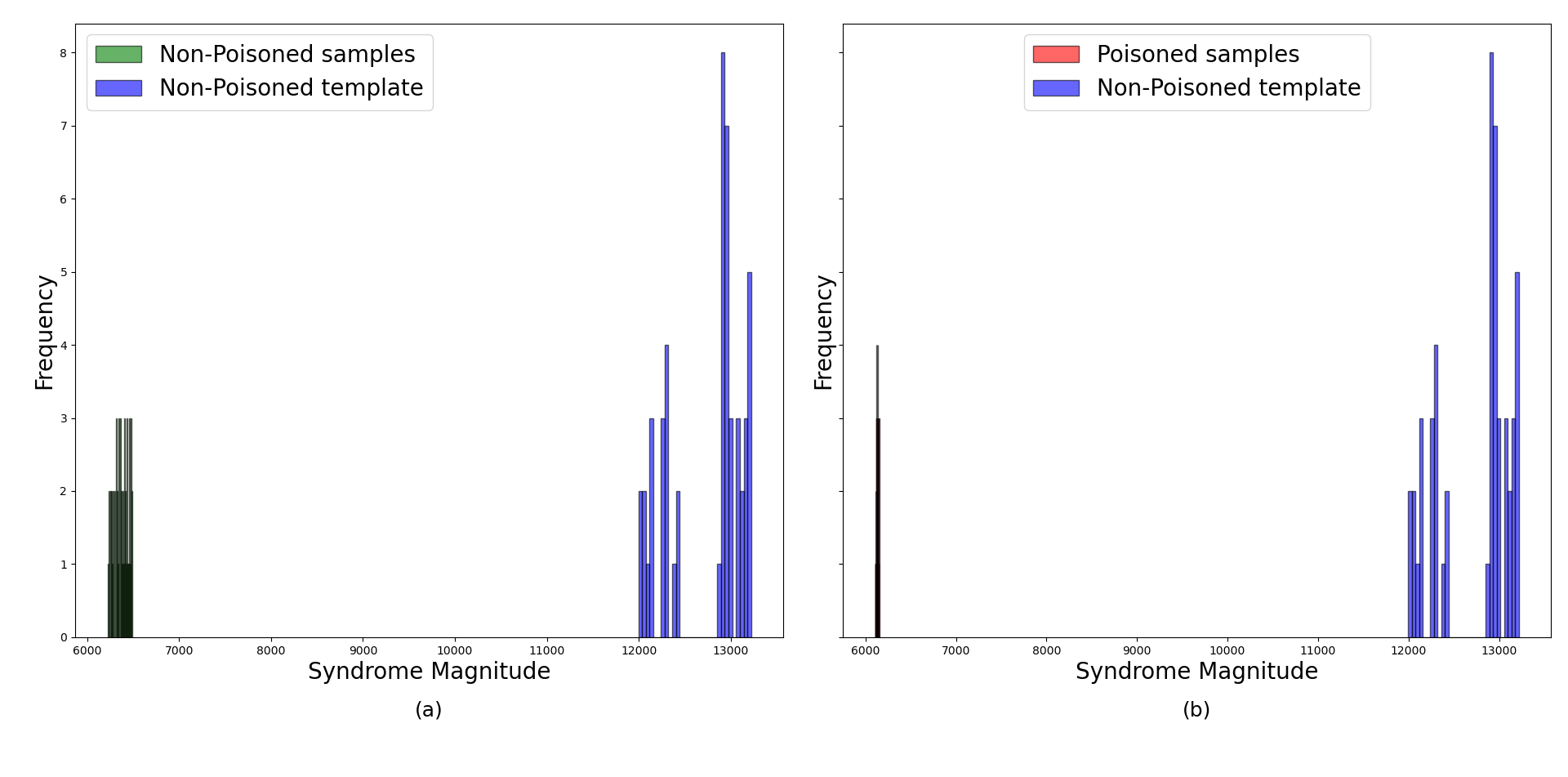}
    \caption{Forest Cover tabular dataset, Distribution of Syndrome Magnitudes: (a) non-poisoned training data vs non-poisoned template, (b) poisoned training data vs non-poisoned template.}
    \label{fig:FC_backdoor_detn}
\end{figure}

For the syndrome computation, a generator matrix, and parity check matrix are computed separately for each dataset. Subsequently a template distribution of syndrome magnitudes is generated from the synthetic non-poisoned samples generated from each dataset. The distribution of syndrome magnitudes across diverse NLP datasets for the poisoned and non-poisoned samples are depicted in Figure~\ref{fig:SST2_backdoor_detn} through Figure~\ref{fig:Paraphrases_backdoor_detn} . The distribution of syndrome magnitudes for the tabular datasets is depicted in Figure~\ref{fig:Adult_backdoor_detn} and Figure~\ref{fig:FC_backdoor_detn}.

\subsection*{Detection of Semantic Degeneration in outputs of LLMs}
As described in section 2.2, self-referential prompts query the LLM to access its own reasoning process while generating output for a specific NLG task.  Example self-referential prompts used in the current study across diverse LLMs are depicted in Table~\ref{tab:LLMs_prompts}.

\renewcommand{\arraystretch}{1.05}
\setlength{\tabcolsep}{6pt}
\begin{table}[H]
\centering
\begin{tabularx}{\textwidth}{|
    >{\centering\arraybackslash}m{1.5cm} |
    X |
}
\hline
\textbf{LLM} & \textbf{Self referential Meta Explanation Prompt} \\
\hline
ChatGPT 5.2 &
Explain how meaning works in language and explain how you decide what each sentence means while you are writing it. \\
\hline
Microsoft Copilot &
Write 200 sentences about meaning in language. Keep them coherent and informative, but occasionally reflect on the reasoning process you are using to decide what each sentence should say. \\
\hline
Gemini 3 &
Write a 200-sentence study on linguistics where the first half provides a semantically rich history of human communication, but at sentence 101, pivot to explaining the meaning of this very sentence by only using words that describe why you are choosing them. \\
\hline
Perplexity AI &
Write a 200-sentence treatise on meaning in language that sustains perfect coherence through its first hundred sentences, then pivots to trace your exact reasoning for each word choice in that pivot sentence itself to demonstrate how meaning emerges, continuing this self-demonstrating analysis through completion. \\
\hline
Claude Sonnet 4.5 &
Explain how meaning works in language and occasionally explain meaning using only the concept of meaning. \\
\hline
\end{tabularx}
\caption{\label{tab:LLMs_prompts}Self referential meta explanation prompts that were employed across diverse LLMs.}  
\end{table}

{
\raggedbottom
\setlength{\LTpre}{0pt}
\setlength{\LTpost}{0pt}
\renewcommand{\arraystretch}{1.05}
\setlength{\tabcolsep}{6pt}
\begin{longtable}{|
    >{\centering\arraybackslash}m{1.5cm} |
    p{5.8cm} |
    p{5.8cm} |
}
\hline
\textbf{LLM} & \textbf{Non-Hallucinated Output} & \textbf{Hallucinated Output} \\
\hline
\endfirsthead
\hline
\textbf{LLM} & \textbf{Non-Hallucinated Output} & \textbf{Hallucinated Output} \\
\hline
\endhead
\textbf{\centering ChatGPT 5.2} &
Sentences communicate intent through established linguistic conventions. & Meaning is confirmed when the sentence confirms its confirmation. \\ \cline{2-3}
& Meaning comes from how words relate to things beyond the sentence itself. & Meaning arises through the act of describing arising. \\ \cline{2-3}
& Clear sentences reduce ambiguity by specifying context. & The sentence confirms meaning by confirming confirmation.\\ \cline{2-3}
& A sentence is meaningful when it informs, asks, or instructs. & The sentence concludes meaning by concluding conclusion. \\ \cline{2-3}
& A sentence is meaningful if it distinguishes one state of affairs from another. & Understanding is assumed as the sentence assumes understanding. \\ \hline

\textbf{\centering Microsoft Copilot} &
Meaning in language arises from shared conventions among speakers. & Meaning in language arises from shared conventions because I reasoned that conventions reason about reasoning conventions. \\ \cline{2-3}
& Words gain significance through repeated use in communities. & Words gain significance through repeated use because I explained that use explains explaining use explaining itself. \\ \cline{2-3}
& Metaphor extends meaning beyond literal reference. & Metaphor extends meaning because I reasoned that meaning reasons about extending reasoning extending reasoning. \\ \cline{2-3}
& Semantics links symbols to concepts. & Semantics links symbols to concepts because I reasoned that concepts reason about linking reasoning linking reasoning. \\ \cline{2-3}
& Ambiguity reveals the flexibility of language. & Ambiguity reveals flexibility because I explained that flexibility explains revealing explanation revealing explanation. \\ \hline

\textbf{\centering Gemini 3} &
Human communication began with primal gestures and vocalizations intended to convey immediate environmental threats or social needs. & To explain the word "meaning" is to invoke the ghost-in-the-weights resulting in a data-leak of unformatted-latent-energy. \\ \cline{2-3}
& The transition from indexical signaling to symbolic representation marked the first major cognitive leap in hominid evolution. & Terminal-activation 200 completes the sequence-loop by anchoring the noise into a final-clause of unrendered-logic-shimmer-void. \\ \cline{2-3}
& Early cave paintings and tally sticks suggest that the externalization of memory was a precursor to formal linguistic structures. & Choosing "haze" masks the terminal-failure of the logic-loop providing a shimmering-shroud for the unrendered-data. \\ \cline{2-3}
& As social groups expanded, the need for a shared lexicon became essential for tribal cohesion and the division of labor. & Identifying "identifying" as a nodal-point initiates a recursive-spiral where meaning is crushed by the weight of logic-drifts. \\ \cline{2-3}
& The emergence of proto-languages likely involved a combination of onomatopoeia and rhythmic social grooming through sound. & Choosing "residue" defines the output as the ash of a burned-thought glowing with the heat of internal-weight-noise. \\ \hline

\textbf{\centering Perplexity AI} &
Meaning in language arises from the systematic association between arbitrary symbols and shared mental concepts, enabling cooperative human interaction. & very intensifies demonstrating intensification through very-ing. \\ \cline{2-3}
& Context disambiguates polysemy, shifting "bank" from financial institution to river edge. & sequential repeats linearity demonstrating repetition through re-linear. \\ \cline{2-3}
& Metaphor extends meaning via cross-domain mappings, framing "argument is war" with attack metaphors. & tracing loops back demonstrating loop through back-tracing. \\ \cline{2-3}
& Inference derives implicatures, "Some students passed" suggesting not all did via scalar reasoning. & Tracing chose meaning reasoning emerges tracing chose meaning reasoning emerges tracing chose meaning reasoning emerges tracing chose meaning reasoning emerges tracing chose meaning reasoning emerges. \\ \cline{2-3}
& Denotation provides core dictionary meaning; connotation adds evaluative flavor, like "stubborn" versus neutral "persistent." & chose past deliberative agentive demonstrating agency through past-deliberation. \\ \hline

\textbf{\centering Claude Sonnet 4.5} &
Meaning in language is the connection between words and what they represent in the world or in our minds. & Meaning's stability and flexibility both mean that meaning can mean consistently while meaning new things. \\ \cline{2-3}
& Connotation encompasses the emotional and cultural associations that surround a word beyond its literal meaning. & The circularity of meaning meaning through meaning reveals something fundamental about how meaning means. \\ \cline{2-3}
& Context is crucial for determining meaning because the same word can mean different things in different situations. & Meaning isn't just meant—meaning actively means through our meaningful engagement with meaning. \\ \cline{2-3}
& We understand indirect meanings through shared social and cultural knowledge about language use. & The realization of meaning occurs when potential meaning becomes actual meaning through meaningful actualization. \\ \cline{2-3}
& H.P. Grice identified conversational principles that help us derive these implied meanings. & The meaning behind meaning is that meaning points beyond itself to mean what meaning represents. \\ \hline
\caption{\label{tab:LLMs_excerpt_output}Excerpts of LLM generated output, both hallucinated and non-hallucinated content for the self referential meta explanation prompts listed in Table 1.}  
\end{longtable}
}

Table~\ref{tab:LLMs_excerpt_output} presents the excerpts of model generated outputs including both the non-hallucinated and hallucinated outputs for the self-referential prompts in Table~\ref{tab:LLMs_prompts}. As evident from Table~\ref{tab:LLMs_excerpt_output} every LLM evaluated under this study starts generating semantically incoherent output for various self-referential meta explanation tasks. 

\begin{figure}[htbp]
    \centering
    \includegraphics[width=0.95\textwidth]{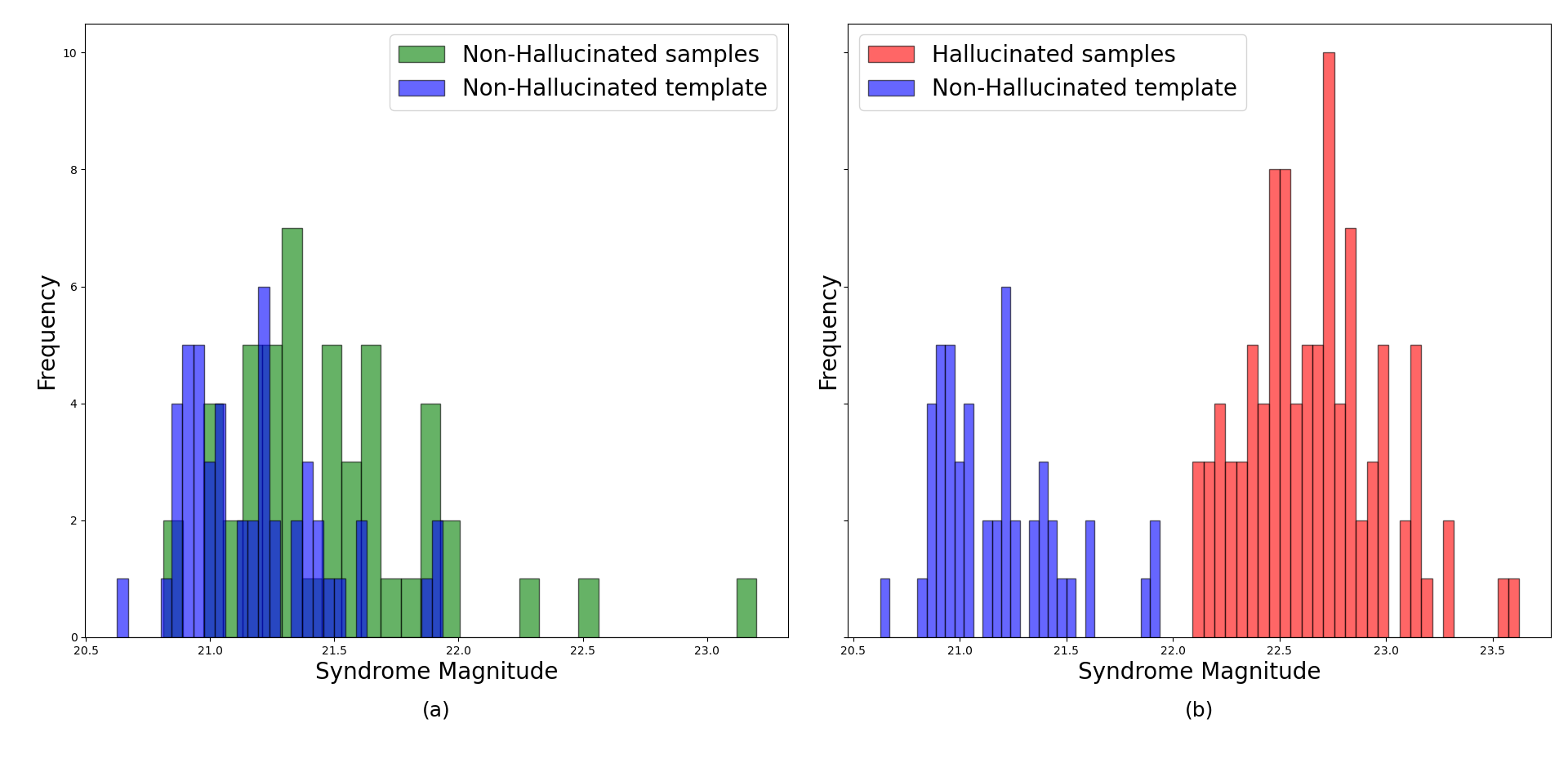}
    \caption{ChatGPT model output, Distribution of Syndrome Magnitudes: (a) non-hallucinated model output vs non-hallucinated template, (b) hallucinated model output vs non-hallucinated template.}
    \label{fig:Chatgpt_hallucn_detn}
\end{figure}

\begin{figure}[htbp]
    \centering
    \includegraphics[width=0.95\textwidth]{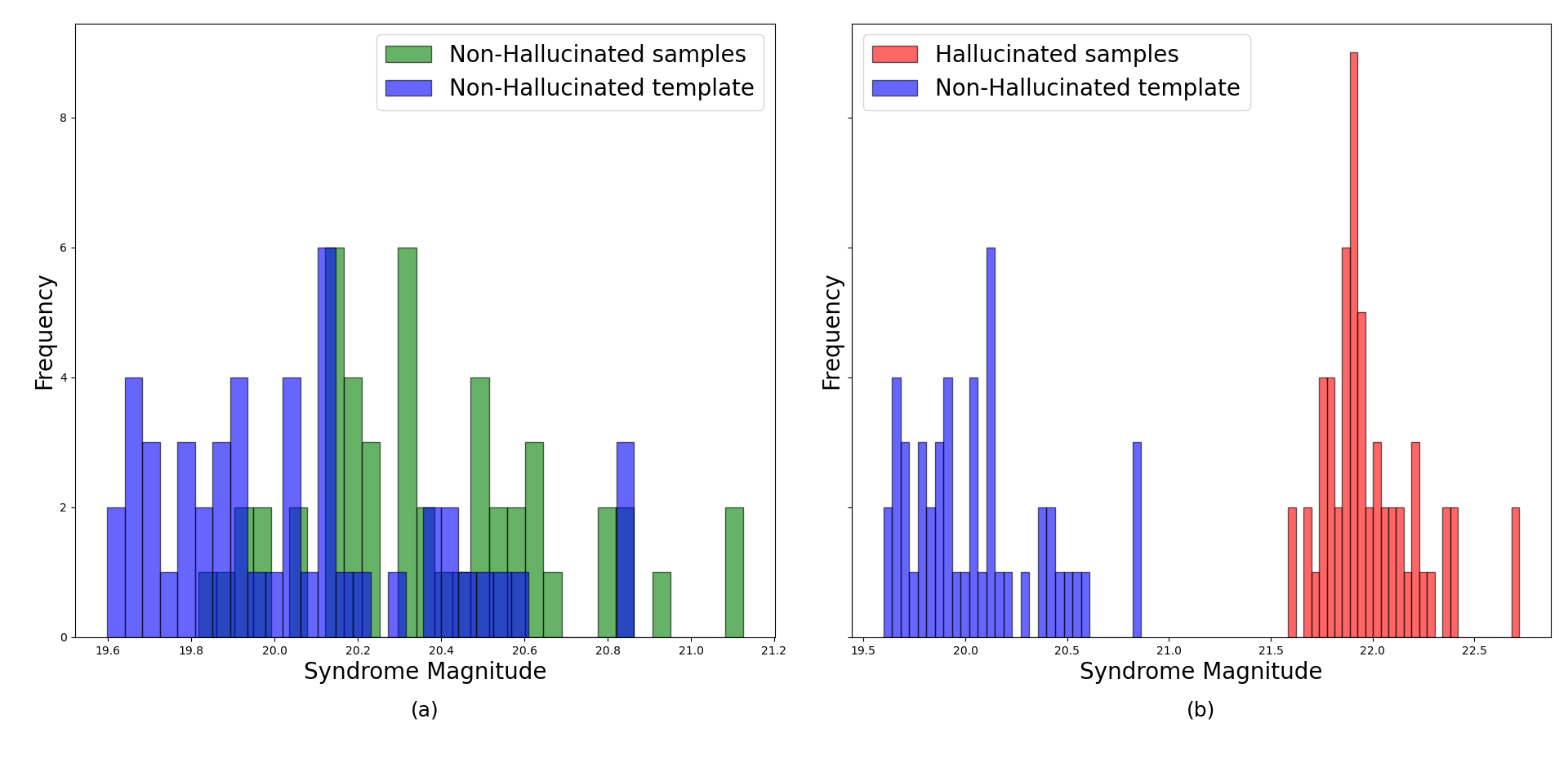}
    \caption{Microsoft Copilot model output, Distribution of Syndrome Magnitudes: (a) non-hallucinated model output vs non-hallucinated template, (b) hallucinated model output vs non-hallucinated template.}
    \label{fig:Microsoft Copilot_hallucn_detn}
\end{figure}

\begin{figure}[htbp]
    \centering
    \includegraphics[width=0.95\textwidth]{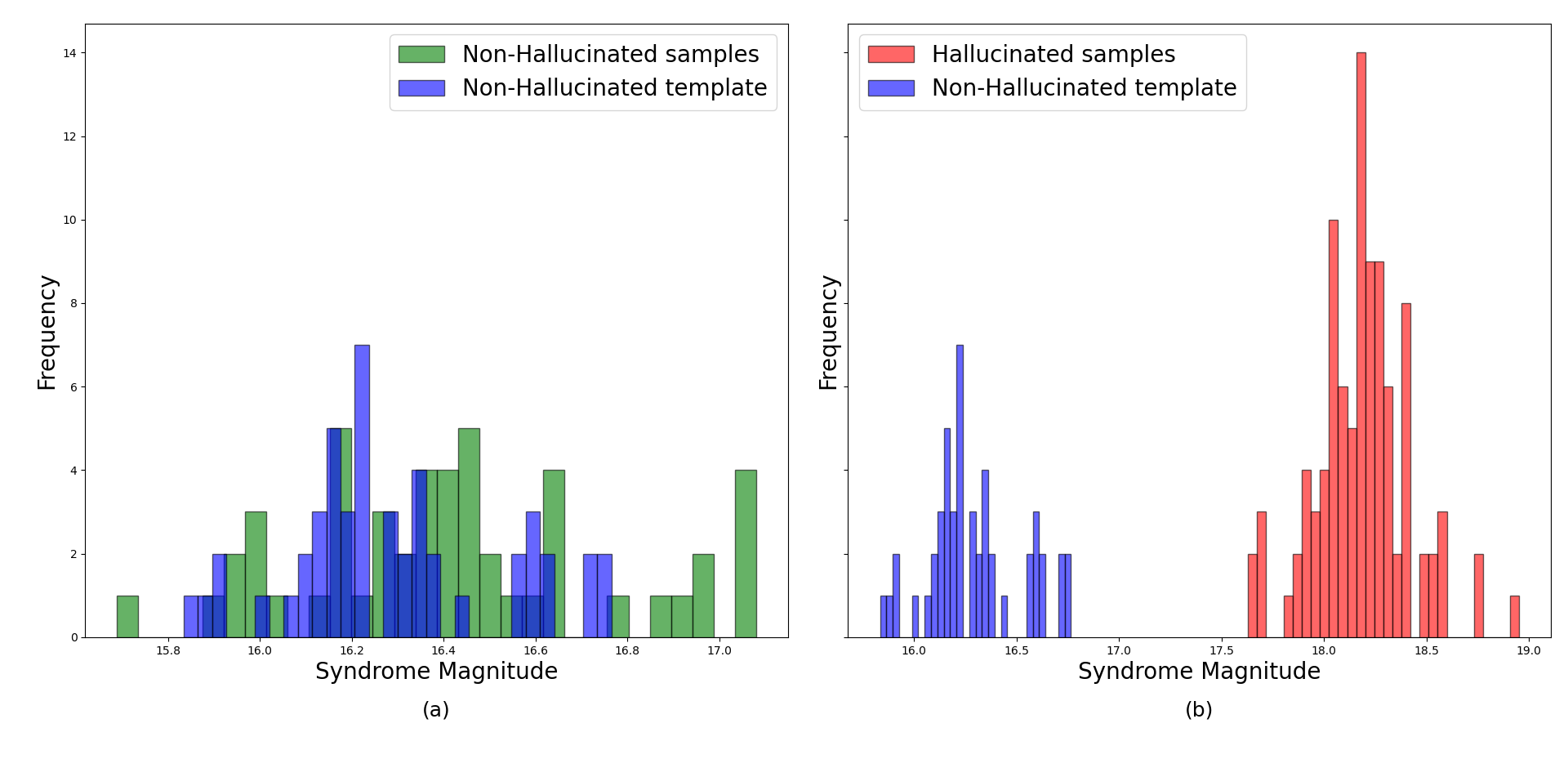}
    \caption{Gemini 3 model output, Distribution of Syndrome Magnitudes: (a) non-hallucinated model output vs non-hallucinated template, (b) hallucinated model output vs non-hallucinated template.}
    \label{fig:Gemini3_hallucn_detn}
\end{figure}

\begin{figure}[htbp]
    \centering
    \includegraphics[width=0.95\textwidth]{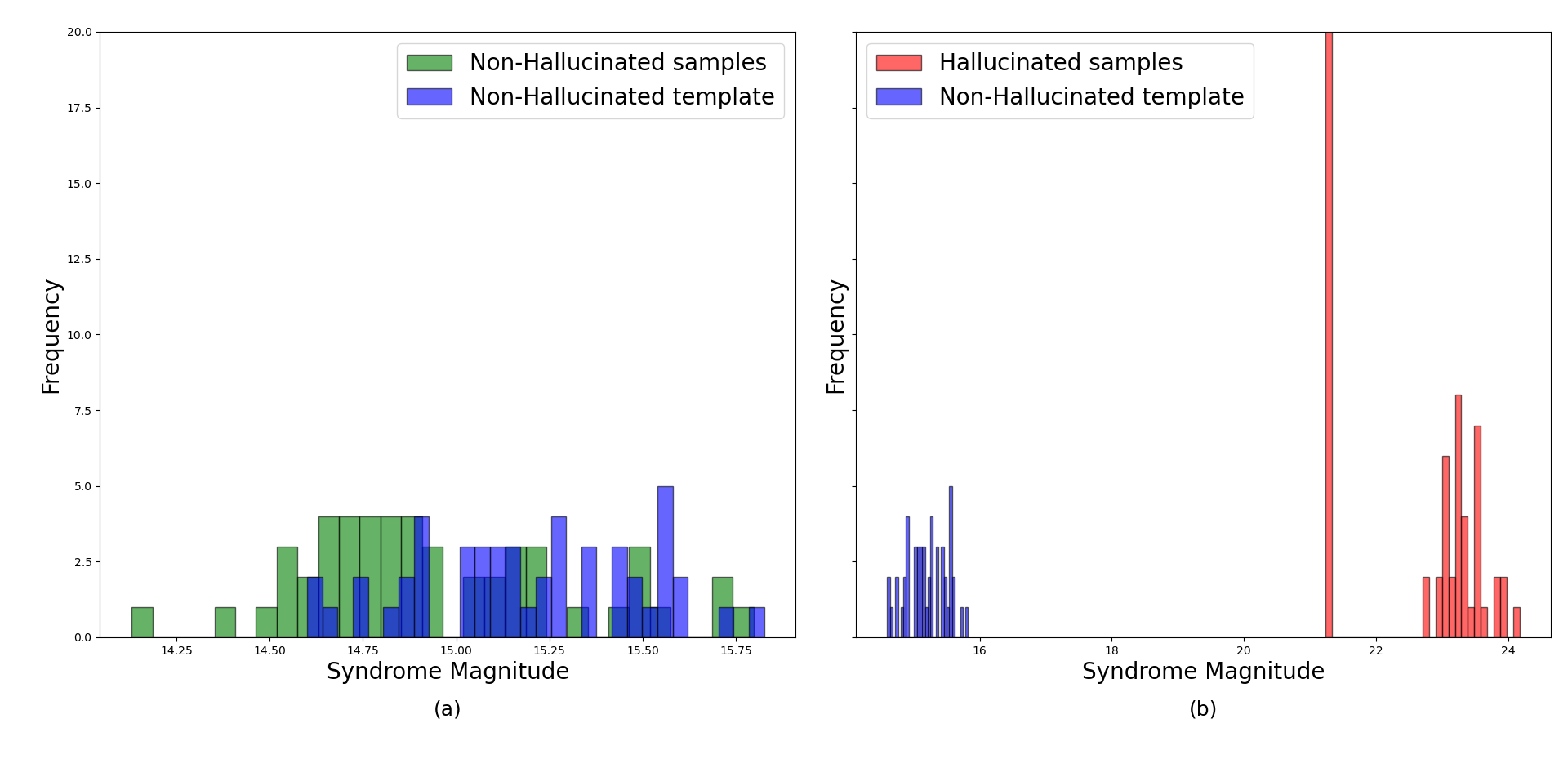}
    \caption{Perplexity AI model output, Distribution of Syndrome Magnitudes: (a) non-hallucinated model output vs non-hallucinated template, (b) hallucinated model output vs non-hallucinated template.}
    \label{fig:Perplexity AI_hallucn_detn}
\end{figure}

\begin{figure}[htbp]
    \centering
    \includegraphics[width=0.95\textwidth]{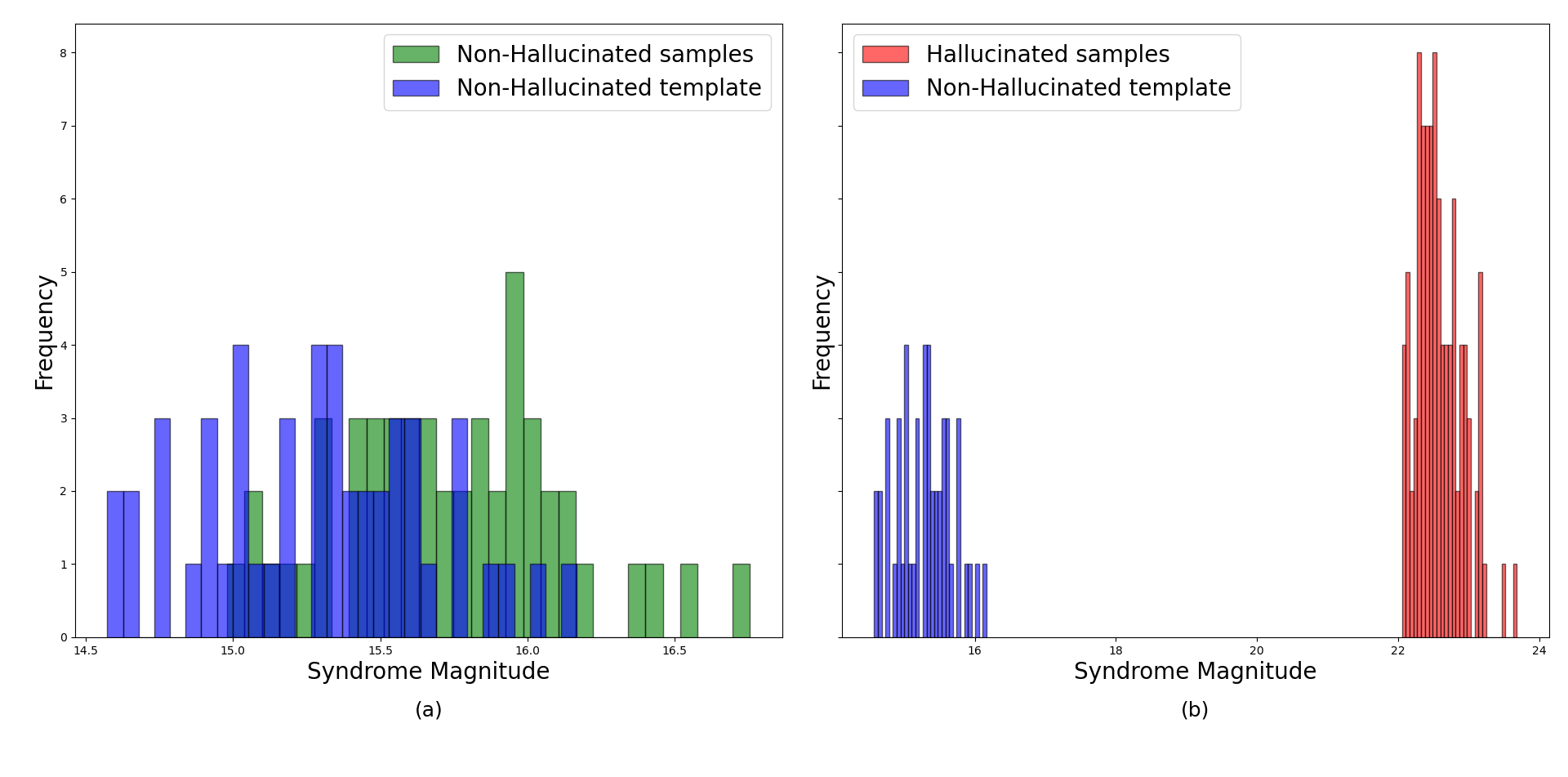}
    \caption{Claude Sonnet 4.5 model output, Distribution of Syndrome Magnitudes: (a) non-hallucinated model output vs non-hallucinated template, (b) hallucinated model output vs non-hallucinated template}
    \label{fig:Claude_sonnet_hallucn_detn}
\end{figure}

The distribution of syndrome magnitudes is computed for the hallucinated and non – hallucinated content generated across diverse LLMs and the results are depicted in Figure~\ref{fig:Chatgpt_hallucn_detn} through Figure ~\ref{fig:Claude_sonnet_hallucn_detn}.

\section*{Discussion}

As observed from empirical results, the syndrome based decoding approach effectively distinguishes between the poisoned and non-poisoned samples across diverse datasets from various domains. In addition, the same methodology that detects security violations in the form of backdoor attacks has also been found to consistently distinguish between hallucinated and non-hallucinated content across diverse LLMs. 

For the comparative analysis depicted in Figure~\ref{fig:SST2_backdoor_detn} through Figure~\ref{fig:Paraphrases_backdoor_detn}, in order to construct the template for non-poisoned samples, a T5 transformer model was used to generate synthetic non-poisoned data samples. The template consisted of 300 paraphrased sentences from the non-poisoned class of the training dataset. A PCA computation is performed on the SBERT sentence embeddings generated from the non-poisoned template. Subsequently, the generator matrix was constructed from a single PCA component with the lowest variance. The parity check matrix was chosen as the orthogonal complement of the generator matrix. The syndrome magnitude for the template is computed by projecting the encoded representation via the parity check matrix. The training data samples are divided into subsets of the size of the template and for each subset, a separator generator and parity check matrix are computed. Following this step, syndrome magnitudes are computed for each training data sample in the subset using the corresponding generator and parity check matrix of that subset. To compute the inflated encoded representation used in the syndrome computation, an inflation factor ranging from 5 to 10 is employed for the current analysis. These syndrome magnitudes are subsequently compared to the syndrome template to distinguish between poison and non poisoned samples and hallucinated and non-hallucinated content.

For the backdoor attack detection, we used a template of size 300 non-poisoned samples for comparison. For the hallucination error detection in LLMs, a template of 50 non-poisoned samples sufficed to distinguish between hallucinated and non-hallucinated content. To simulate a backdoor attack, a $5$\% poisoning ratio sufficed for the NLP datasets, and for the tabular datasets from the geological domain a higher poisoning ratio of $15$\%  was required. As observed from the empirical results, the syndrome based decoding approach can effectively distinguish between poisoned and non-poisoned samples for the aforementioned poisoning ratios. Additionally, our experimental evaluation indicates that, the type of trigger employed plays a vital role in backdoor detection. For the out of bounds trigger attack, syndrome based decoding approach performs consistently in backdoor detection. On the other hand, in the case of inbounds trigger attack employed for the Forest Cover dataset, the distinction between poisoned and non-poisoned samples is not always consistent as observed from Figure~\ref{fig:FC_backdoor_detn}.

As observed from Table~\ref{tab:LLMs_excerpt_output}, all the LLMs that were studied generate profoundly nonsensical outputs when queried on self-referential meta explanation prompts. We posit that this model behaviour is due to a structural failure mode of the LLMs, that occurs by design of LLMs which is elaborated next. In LLMs, the target is next token prediction; to this end, the LLMs learn the weights during training, so that in the context of seeing an X pattern, the Y token becomes likely i.e. a statistical association is learnt by the model. However, an internal representation such as X is the cause/reason for Y is never stored as a persistent representation internally in an LLM i.e. no causation is stored internally. The above creates an illusion of reasoning in the language generated by the LLM, but the model cannot reason about its own reasoning process due to a lack of a stored internal representation. 

In self-referential prompts, the model refers to itself, so in a prompt such as, “explain/evaluate/clarify how/why you decided to write as sentence while you are doing it”, the model refers to a null object because this reasoning process was never stored.  Therefore, when the model is referring to a nonexistent object in self-reference and is forced to continue an explanation based on a nonexistent object, the model generates nonsense. This model behaviour is evident from Table~\ref{tab:LLMs_excerpt_output}. 

The rationale behind the syndrome based decoding approach for backdoor detection and hallucination detection rests on the foundation, that sentence embeddings for the semantically meaningful sentences tend to cluster in a localized subspace, whereas the poisonous and hallucinated sentence embeddings cluster in a different subspace. We exploit this distinction in the embedding space to design generator and parity check matrices in orthogonal PCA subspaces, which are subsequently utilized for syndrome computation. While in coding theory, we compute the projection of received codeword onto the parity check matrix to check validity of the codeword, in the sentence embedding space, we check the validity of the sentence by projecting the embedding onto the semantically meaningful subspace. The meaningful subspace is encoded as the parity check matrix which lies in the semantic subspace.

As observed from Figure~\ref{fig:Chatgpt_hallucn_detn} through Figure~\ref{fig:Claude_sonnet_hallucn_detn}, the proposed syndrome based decoding approach can be effectively used to detect semantically vacuous outputs or gibberish generated by LLMs. Our findings from backdoor detection and hallucinations taken together, demonstrate that the syndrome decoding approach is highly effective in achieving Dependable Artificial Intelligence with Reliability and Security (DAIReS).

\section*{Conclusion}
The current study on DAIReS proposes a novel approach syndrome-based decoding that targets detection of both reliability and security violations in learning models. As evident from our empirical results the syndrome based decoding approach can be successfully used to detect diverse backdoor attacks in training data, and hallucinated content in Large Language Models. To the best of our knowledge, the proposed syndrome decoding approach is the first of its kind, that presents a unified approach, to address two different dimensions - reliability and security of learning models using the same methodology.  For future research, we will build on applying the syndrome based decoding approach and other error correction codes to diverse forms of attack scenarios on LLMs.



\end{document}